\documentclass[11pt,a4paper]{article}
\usepackage{amsfonts}
\usepackage{jheppub}

\title{Non-minimal Coupling Branes}

\author{Yu-Xiao Liu, Feng-Wei Chen\footnote{Corresponding author},
        Heng Guo, Xiang-Nan Zhou}

\affiliation{Institute of Theoretical Physics, Lanzhou University,\\
Lanzhou 730000, People' s Republic of China}

\emailAdd{liuyx@lzu.edu.cn, chenfw10@lzu.edu.cn, guoh2009@lzu.cn,
zhouxn10@lzu.edu.cn}

\abstract{ We study a thick and $\mathbb{Z}_2$ symmetric braneworld model with a non-minimally coupled bulk scalar.
Several analytic solutions are found. There are
two categories of the brane configuration: standard branes and deformed ones. The
former is just the same with the solutions in general relativity, whereas the latter
has negative effective energy densities in the location of the branes. The
question of stability under metric tensor perturbation is also investigated, and there is
no tachyon. We show that the gravity zero mode (namely the 4D massless graviton)
is the only one state that can be localized on the branes. So Newtonian potential can
be recovered in 4D space-time. }

\begin{document}

\maketitle

\section{Introduction}

In the 60's of 20th century, Brans and Dicke proposed that in gravity theory there
should exist a scalar field according to Mach's principle. From then on, both
Brans-Dicke theory and general relativity are generally held to be in agreement with
observations. The Brans-Dicke theory has been generalized to the so-called
scalar-tensor theory of gravity (see a comprehensive book~\cite{FujiiMaed2003}). In
string/M theories, there is also a scalar field (dilaton), which is non-minimally
coupled to gravity. Therefore, string/M theories give the scalar-tensor
theory\footnote{In this paper, we ignore the difference between the scalar-tensor
theory and the non-minimal theory of gravity.} rather than general relativity. It
has been shown that various $F(R)$ theories are also equivalent to non-minimal coupling
theories (see e.g. ~\cite{DeTsujikawa2010}). Furthermore, non-minimally
coupled theories can also originate from multi-dimensional theories, quantum field
theory in curved space-time, and induced gravity theories.

In a general gravity theory with a non-minimally coupled scalar, higher derivative terms can also
be included. The action is written as
 \begin{equation}
 S=\int_M \sqrt{-g}d^n x \left[\mathfrak{L}_{g}(f(\phi,X)R, R^{AB}\partial_A\phi\partial_B\phi
 ,\cdots)+ \mathfrak{L}_{m}(X,\phi)\right],
 \end{equation}
where $X=-g^{AB}\partial_A\phi\partial_B\phi$. However, the general action is difficult
to deal with since the corresponding equations of
motion are higher derivative equations. In this paper, we would like to investigate the
 non-minimal coupling thick braneworld model in
5-dimensional space-time, and the action is chosen as
 \begin{equation}
 S=\int_M \sqrt{-g}d^5 x \left[\frac{F(\phi)}{16\pi G_5}R+ \mathfrak{L}_{m}(X,\phi)\right].
 \end{equation}
This action does not lead to more complicated equations of motion in principle. We
do not cost too much to solve the problem. If $F(\phi)=e^{\gamma\phi}$, the scalar
is a dilaton. This model has a close relation with Weyl geometry, which has been
extensively studied.

On the other hand, the idea that we live on a brane \cite{RubakovShaposhnikov1983,
Arkani-HamedDimopoulosDvaliMarch-Russell2002, RandallSundrum1999a,
RandallSundrum1999} has received a lot of attention in recent years. Braneworld
models have revealed new possibilities for addressing the hierarchy problem of
particle physics as well as cosmological constant problem. Thin braneworld models,
such as the Arkani-Hamed-Dimopoulos-Dvali (ADD)
model~\cite{Arkani-HamedDimopoulosDvaliMarch-Russell2002} and the Randall-Sundrum
(RS) models ~\cite{RandallSundrum1999a, RandallSundrum1999}, were inspired by D-brane
theory. While, thick branes, which are usually generated by scalar fields, were
proposed as a generalization of thin branes ~\cite{DeWolfeFreedmanGubserKarch2000,
BronnikovMeierovich2003a, BazeiaGomesLosanoMenezes2009, Toharia:2010ex,
Alencar:2010mi},  (see a recent review~\cite{DzhunushalievFolomeevMinamitsuji2010}
 and references therein). In thick
braneworld model, the Standard Model (SM) matter fields are confined dynamically on
the brane.


It is natural to consider braneworld models with non-minimally coupled background scalar
\cite{Bogdanos:2006qw, FarakosKoutsoumbasPasipoularides2007, Farakos:2005hz,
Farakos:2006tt, Farakos:2006sr, Bertolami:2007dt, Andrianov:2007tf, Bogdanos:2006dt,
Setare:2008mb, Guo:2011wr, Granda:2010hb, Koyama:2006mh, Mikhailov:2006vx},
{since braneworld models were inspired by string theories. Several
exact RS metric solutions were obtained in \cite{Farakos:2005hz, Farakos:2006tt,
Bogdanos:2006qw}}. The
models with a special form $F(\phi)=1-\alpha \phi^2$ have been studied in various
papers \cite{Bogdanos:2006qw, FarakosKoutsoumbasPasipoularides2007, Farakos:2005hz,
Farakos:2006tt, Farakos:2006sr, Bertolami:2007dt, Andrianov:2007tf, Bogdanos:2006dt,
Setare:2008mb, Guo:2011wr}. Several exact RS metric solutions were obtained in
\cite{Farakos:2005hz, Farakos:2006tt, Bogdanos:2006qw}. Some thick brane solutions
were also given in \cite{Bogdanos:2006qw}. The accurate perturbation equation was
given in \cite{FarakosKoutsoumbasPasipoularides2007}. In \cite{FarakosKoutsoumbasPasipoularides2007}, the effect
Newtonian potential for RS metric branes was {calculated} by the technic of the bent
brane \cite{Giddings:2000mu, PhysRevLett.84.2778}. Moreover, a non-minimally coupled
Phantom bulk field was considered in \cite{Setare:2008mb}.

Whereas, all results were based on the special form, and no general information was
given in the previous papers. We cannot finger out whether those results are
dependent on the special {choice}. For other {functions} $F(\phi)$, no exact solution is
provided. Therefore, in the present article a general non-minimal coupling
$F(\phi)$ is taken into account. We consider a static $\mathbb{Z}_2$ symmetric flat
3-brane embedded in 5D space-time and an ordinary kink scalar field is non-minimally
coupled to gravity.

In the following section, we introduce the model and fundamental properties of
$F(\phi)R$ gravity. In section \ref{secSolutions}, some exact solutions are given,
and the behavior of the asymptotic space-time is also studied. In section
\ref{secStability}, we investigate the problem of stability. As expected, there is
no tachyon for the gravitational perturbation. We list our main results in the last
section.

\section{The model}
\label{secModel}

In this paper, we consider a five-dimensional gravity theory with a
non-minimally coupled bulk scalar field. The action takes the
following form
\begin{equation}
S=\int d^4x dy\sqrt{-g}\left(\frac{1}{2
}F(\phi)R-\frac{1}{2}(\partial\phi)^2-V(\phi)\right)+\int d^4x
\sqrt{-g^{(\text{brane})}}\lambda(\phi), \label{Action1}
\end{equation}
where $(\partial\phi)^2=g^{AB}\partial_A\phi\partial_B\phi$. For convenience we set
$8\pi G_5=M_5^{-3}=1$, $ x^5= y$. The kinematics of the scalar field is standard. If
$F(\phi)$ is positive, it describes gravity; negative, anti-gravity. If there exists
anti-gravity, there would be forces of repulsion between matter fields in this region. This leads to the result
that matter distribution may be unstable. So we assume that $F(\phi)$ is a general smooth
non-negative function and its zero points lie only at infinity if they exist.
This is a restriction of the scalar-tensor theory. Furthermore, at the location of
the brane $F=1$ is satisfied. Thin brane solutions have been studied in
\cite{Bogdanos:2006qw} \cite{FarakosKoutsoumbasPasipoularides2007}. In the following
we consider thick brane scenario, so the four-dimensional part in (\ref{Action1})
vanishes.

By setting $\Phi=F(\phi)$, we can write the action (\ref{Action1}) as
\begin{equation}
S=\int d^4x dy \sqrt{-g}\left[\frac{1}{2}\Phi R-\frac{\omega(\Phi)}{\Phi}(\partial
\Phi)^2-\tilde{V}(\Phi)\right],
\end{equation}
where the Brans-Dicke parameter $\omega(\Phi)=\frac{F}{2F_\phi^{2}}$
with $F_\phi \equiv \partial_{\phi} F(\phi)$.


We have two ways to solve the field equations of the scalar-tensor gravity theory.  The first one is to make a
conformal transformation to recover Einstein frame:
\begin{eqnarray}
\tilde{g}_{AB}&=&e^{2\omega}g_{AB}=F^{2/3}g_{AB},\label{Einstein_frame_metric}\\
\tilde{R}&=&e^{-2\omega}(R-8\nabla^2\omega-12(\partial
\omega)^2), \\
S &=& \int d^4xdy \sqrt{-\tilde{g}}
  \Bigg[ \frac{1}{2}\tilde{R}
        -\frac{1}{2}\left(\frac{4F_\phi^2}{3F^{2}}
               +\frac{1}{F}
          \right) \tilde{g}^{AB}\partial_A \phi\partial_B \phi
        -F^{5/3}V(\phi)\Bigg].
\end{eqnarray}
Defining a new scalar field
\begin{equation}
\xi=\int
d\phi\sqrt{\frac{4F_\phi^2}{3F^{2}}+\frac{1}{F}},\label{transformation}
\end{equation}
we arrive at
 \begin{equation}
 S=\int d^4xdy
\sqrt{-\tilde{g}} \left[\frac{1}{2}\tilde{R}-\frac{1}{2}\tilde{g}^{AB}\partial_A
\xi\partial_B \xi -\tilde{V}(\xi)\right]
 \end{equation}
with $\tilde{V}=F^{5/3}(\phi(\xi))\;V(\phi(\xi))$. Thus, the problem in the
scalar-tensor gravity is transferred as solving the field equations in Einstein frame and the
transformation (\ref{transformation}).

The second way, which is the one we will adopt in this paper, is to solve the
following modified Einstein equations straightly
 \begin{equation}
F(\phi)\big(R_{AB}-\frac{1}{2}g_{AB}R\big)+( g_{AB}\nabla^2
-\nabla_A\nabla_B)F(\phi)=T^{(\phi)}_{AB}, \label{EinsteinEqa}
\end{equation}
where $T^{(\phi)}_{AB}=\partial_A
\phi\partial_B\phi-g_{AB}\left(\frac{1}{2}(\partial\phi)^2+V\right)$ is the
energy-momentum tensor of the scalar field. We can write them in a standard form:
\begin{equation}
R_{AB}-\frac{1}{2}g_{AB}R=T_{AB}^{\text{eff}} \label{EinsteinEqb}
\end{equation}
with the effective energy-momentum tensor $T_{AB}^{\text{eff}}=\big[T^{(\phi)}_{AB}-(
g_{AB}\nabla^2 -\nabla_A\nabla_B)F(\phi)\big]/F(\phi)$.

In the following, we use another version of the modified Einstein equations (\ref{EinsteinEqa}):
 \begin{eqnarray}
 F(\phi)R_{AB}
    -(\nabla_A\nabla_B+\frac{1}{3}g_{AB}\nabla^2)F(\phi)
    = \bar{T}^{(\phi)}_{AB}, \label{EinsteinEqc}
 \end{eqnarray}
where
\begin{eqnarray}
 \bar{T}^{(\phi)}_{AB}
    = T^{(\phi)}_{AB}-\frac{1}{3}g_{AB} T^{(\phi)}
    = \partial_A\phi\partial_B\phi+\frac{2}{3}g_{AB}V(\phi).
 \end{eqnarray}
The equation of motion for the scalar field is
\begin{equation}
\nabla^2\phi+ \frac{1}{2}F_\phi(\phi)R-V_\phi(\phi)=0.\label{EOMscalar1}
\end{equation}

The metric of a thick brane model is usually assumed as follows \footnote{$A, B$ run
for all 5D indices, $\mu, \nu$ for 4D ones.}
\begin{eqnarray}
  ds^2=g_{AB}dx^Adx^B=e^{2A(y)}\hat{g}_{\mu\nu}(x)dx^\mu dx^\nu+dy^2,
\end{eqnarray}
where $\hat{g}_{\mu\nu}(x)$ is the induced metric on the brane. The scalar curvature
is
\begin{eqnarray}
  R=e^{-2A(y)}\hat{R}-8A''(y)-20A'^2(y),
\end{eqnarray}
where $\hat{R}$ is the Ricci scalar calculated by the induced metric $\hat{g}_{\mu\nu}$. The four-dimensional
Newton's gravity constant $G_4$ is obtained by the dimensional reduction (here we
recover the 5-dimensional Newton's constant)
 \begin{equation}
 \frac{1}{8\pi G_4}=M_p^2=\frac{1}{8\pi G_5}\int^{+\infty}_{-\infty}dy
 e^{2A(y)}F(\phi(y)).
 \end{equation}

In order to construct a static flat thick brane, we use the following metric
\begin{equation}
ds^2=e^{2A(y)}\eta_{\mu\nu}dx^\mu dx^\nu+dy^2,
\end{equation}
which has four-dimensional Poincar\'{e} symmetry. Then the modified Einstein equations
(\ref{EinsteinEqc}) read
 \begin{eqnarray}
 3F(\phi(y))(4A'^2+A'')+7A'F'(\phi(y))+F''(\phi(y))&=&-2V,\label{Einstein1}\\
 3F(\phi(y))A''-A'F'(\phi(y))+F''(\phi(y))+\phi'^2&=&0.\label{Einstein2}
 \end{eqnarray}
Here the prime denotes the derivative with respect to $y$. And the equation of
motion for the scalar field (\ref{EOMscalar1}) reads
 \begin{equation}
\phi''+4A'\phi'-2F_\phi(\phi)(2A''+5A'^2)-V_\phi(\phi)=0.\label{EOMscalar2}
\end{equation}
Among the above three equations (\ref{Einstein1})-(\ref{EOMscalar2}), only two are
independent.

In general relativity, BPS solutions can simplify Einstein equations. We can obtain
solutions from the following way. First, we set
 \begin{eqnarray}
 \phi'=W(\phi),\quad
 A'=P(\phi).
 \end{eqnarray}
Then eq.~(\ref{Einstein2}) becomes
 \begin{eqnarray}
 3FP_\phi-PF_\phi+(F_{\phi\phi}+1)W+F_\phi W_\phi=0,
  \end{eqnarray}
and the scalar potential is given by
  \begin{equation}
  V=\frac{1}{2}W^2-6P^2F-4PWF_\phi.\label{V-phi}
  \end{equation}

\section{The solutions}
\label{secSolutions}

In conventional thick brane theory, there is a scalar field with a proper
symmetry-breaking potential as an order parameter. So we often choose the scalar
field as a kink solution. The kink interpolates between two anti-de Sitter vacua, at
which the scalar potential reaches its (local) minimal values. However, there is some
subtle difference for the non-minimally coupled theory, since the term $F(\phi)R$ contributes an
effective potential for the scalar field. We can choose proper functions $F(\phi)$
and $V(\phi)$ to construct various branes.

\subsection{The asymptotic property of the warp factor}

In this paper we mainly study symmetric thick branes generated by a single kink
scalar. So we choose that $\phi(y)$ is an odd function and $A(0)=A'(0)=0$. It imposes
the constraint that $F(\phi)$ and $V(\phi)$ are even functions of $\phi$.

From eq. (\ref{Einstein2}), it is easy to obtain
 \begin{equation}
 A''(0)=-\frac{1}{3}\big[F''(\phi(0))+\phi'(0)^2\big]
       =-\frac{1}{3}\big[F_{\phi\phi}(\phi(0))+1\big]\phi'(0)^2.
 \end{equation}
Here we only consider the single kink scenario, hence $\phi'(0)^2>0$. With the help
of eq. (\ref{Einstein1}), we have the effective energy density:
\begin{eqnarray}
 \rho_{\text{eff}}=-g^{00}T_{00}^{\text{eff}}=\frac{1}{2F}
           \left(\phi'^2+F''-3A''F+3A'F'-12A'^2F\right) .
\end{eqnarray}
If $F_{\phi\phi}(\phi(0))>-1$, then $A''(0)<0$ and we will get a usual single
brane just like in the standard thick brane case. If $F_{\phi\phi}(\phi(0))<-1$,
then $A''(0)>0$ and the brane will be split into two positive tension sub-branes
located around $y=0$ and a negative tension sub-brane located at $y=0$, which cannot
be constructed in conventional general relativity.

As is well known, in Einstein's gravity, a kink solution will result in
an asymptotic AdS space-time. Now we analyze the asymptotic space-time
in the non-minimally coupled theory considered in the paper.

For given $F(\phi)$ and $\phi(y)$, $A'(y)$ can be calculated by
\begin{eqnarray}
A'(y)&=&-\frac{1}{3}F^{\frac{1}{3}}(\phi(y))\int_0^{y}
F^{-\frac{4}{3}}(\phi(\bar{y}))\Big[F''(\phi(\bar{y}))+\phi'^2\Big]d\bar{y}\label{A'1}  \\
&=&-\frac{F'(\phi(y))}{3F(\phi(y))}-\frac{1}{3}F^{\frac{1}{3}}(\phi(y))\int_0^{y}
    \Big[ \frac{4F'^2(\phi(\bar{y}))}
               {3F^{\frac{7}{3}}(\phi(\bar{y}))}
         +\frac{\phi'^2(\bar{y})}
               {F^{\frac{4}{3}}(\phi(\bar{y}))}
    \Big]d\bar{y}.  \label{A'}
\end{eqnarray}
Here we assume that $F(\phi)$ is an even smooth function of $\phi$. If $F(\phi(y))$ tends to infinity
when $y\rightarrow\infty$, the warp factor $e^{2A(y)}<F^{-2/3}(\phi(y))$ and vanishes. If $F(\phi(\infty))$ is {a} positive constant (notice that $F'(\phi(0))=0$), we would have
 \begin{eqnarray}\label{Afull}
 A'(+\infty)=-\frac{1}{3}F^{\frac{1}{3}}(\phi(+\infty))\int_0^{+\infty}
    \Big[ \frac{4F'^2(\phi(\bar{y}))}
               {3F^{\frac{7}{3}}(\phi(\bar{y}))}
         +\frac{\phi'^2(\bar{y})}
               {F^{\frac{4}{3}}(\phi(\bar{y}))}
    \Big]d\bar{y}   .
 \end{eqnarray}
From the above equation we know that $A'(+\infty)$ is a negative constant, so the scalar curvature
tends to a negative constant at infinity, and the space-time is asymptotic
AdS. This is a general conclusion. This conclusion can be confirmed
by the way of the conformal transformation.

In the following, we study the possible category of the asymptotic space-times when
  $F(\phi(\infty))=0$.
  One may expect that there exists asymptotic
  Minkowski space-time for proper $F(\phi)$ and $V(\phi)$, since the contribution of gravity
  could vanish at infinity. While, for a kink solution, the warp factor $A(y)$
  cannot be a finite constant at infinity. Hence, the space-time is asymptotic
  Minkowski in the sense that the scalar curvature is zero at infinity.

The second term of eq.~(\ref{Afull}) makes $A(y)$ descend faster, so in
the following we mainly concentrate on the first term. We assume
$F(\phi(y))$ can be expanded at infinity:
 \begin{equation}
 F\sim a |y|^{-b}, \quad |y|\rightarrow+\infty, \label{Fyinfty11}
 \end{equation}
 where $a>0$, $b>0$, and '$\sim$' denotes that we omit the higher order terms of $(y^{-b})$.

For the case $b>3$, though the integral
$\int_0^{+\infty}dy F^{-\frac{4}{3}}(y)F''(y)$ in (\ref{A'1}) is divergent, we
have the asymptotic form of $A'(y)$ for large $y$
 \begin{equation}
 A'(y)\sim -\frac{1}{3}F^{\frac{1}{3}}(y)\int_0^{y}d\bar{y}
 F^{-\frac{4}{3}}(\bar{y})F''(\bar{y})\sim-\frac{b(b+1)}{b-3}y^{-1}<0,~~y\rightarrow+\infty.
 \end{equation}
This term leads that $A(y)\sim -\frac{b(b+1)}{b-3}\ln |y|$ is negative infinity.

When $b=3$, we can get by simple calculation
 \begin{eqnarray}
 A'(y)&\sim&-\frac{\ln y}{3y},\quad y\rightarrow+\infty,\\
 A(y)&\sim&-(\ln |y|)^2/3,\quad |y| \rightarrow +\infty.
 \end{eqnarray}

For $b<3$, the first term of eq.~(\ref{Afull}) is convergence, but the property of
$A'(y)$ at infinity is also mainly determined by the asymptotic expansion of $F(y)$.
 From eq. (\ref{A'}), we have
 \begin{equation}
 A'(y)\sim\frac{b}{3y}-Cy^{-\frac{b}{3}},\quad y\rightarrow+\infty,
 \end{equation}
 where $C$ is a positive integration constant.
Since $y^{-1}$ descends faster than $y^{-\frac{b}{3}}$,
 \begin{equation}
 A(y)\sim -C(1-\frac{b}{3})|y|^{1-\frac{b}{3}},~~y\rightarrow+\infty.
 \end{equation}
Extending $b\rightarrow0$, we get
 \begin{equation}
 A(y)\sim -C|y|,~~|y|\rightarrow+\infty.
 \end{equation}
This agrees with the conclusion obtained from the situation in which $F(\phi(y))$ is
a positive constant at infinity.

From the above discussion we know that the asymptotic property of $A'(|y|\rightarrow\infty)$ is mainly determined by $F(\phi(y))$
and $ \phi(y)$ at infinity. In all above cases, $A'(|y|\rightarrow\infty)$ approaches to zero, however $A(|y|\rightarrow\infty)$ tends to negative infinity.

We can prove that for $F(\phi(y))\sim e^{-a|y|^b}$ or $F(\phi(y))\sim a|y|^{-b}(\ln
|y|)^c$ when $|y|$ tends to infinity, the warp factor $e^{2A(y)}$ vanishes.  We conjecture that in our brane model  the warp factor $e^{2A(y)}$ should vanish (It could be a positive constant as an extreme result). Nevertheless, we cannot find
a formula to prove this for general functions.

\subsection{The standard solutions}

\begin{figure*}[htb]
\begin{center}
\includegraphics[width=4.5cm]{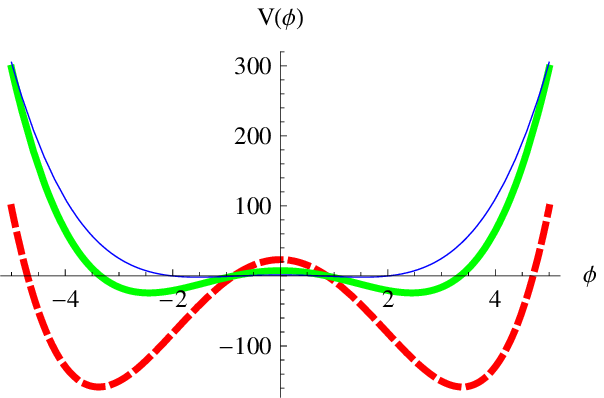}
\includegraphics[width=4.5cm]{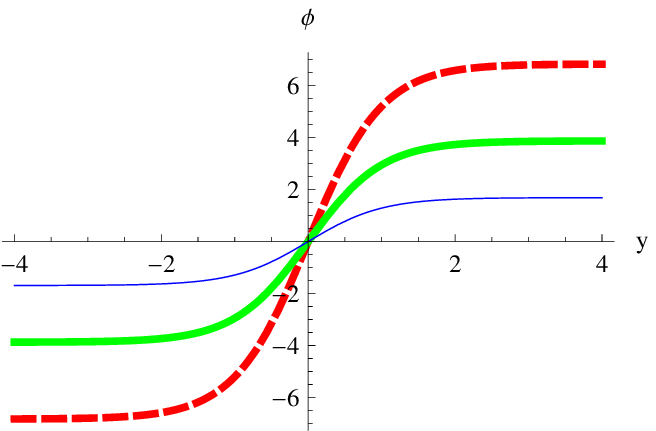}
\includegraphics[width=4.5cm]{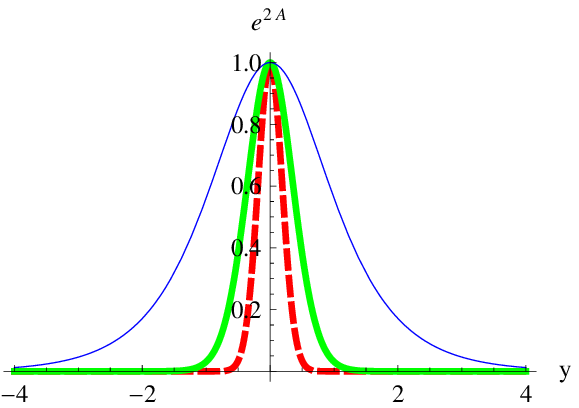}
\end{center}
 \caption{ The shapes of the potential $V(\phi)$, the scalar $\phi$  and the warp factor $e^{2A}$
 for the exact solution 1.
The dashing red, thick green and thin blue lines correspond to $\alpha=0.05, 0.10$
and 0.15, respectively. The parameter $\kappa$ is set as
$\kappa=1$.}\label{figVphiE2A_1}
\end{figure*}

 \textbf{Exact solution 1}:

 A kink solution was given
by Bogdanos, Dimitriadis and  Tamvakis  in \cite{Bogdanos:2006qw}. In this solution,
$F(\phi)=1-\alpha \phi^2$ with $\alpha$ a dimensionless constant, and the scalar and
the warp factor are given by
 \begin{eqnarray}
 \phi&=&\phi_0\tanh\kappa y=\sqrt{\frac{3(1-6\alpha)}{\alpha(1-2\alpha)}}\tanh(\kappa y),\\
 A&=&(\alpha^{-1}-6)\ln\cosh(\kappa y),
 \end{eqnarray}
where $\phi_0=\sqrt{\frac{3(1-6\alpha)}{\alpha(1-2\alpha)}}$.
This solution is only valid for the range $0<\alpha<\frac{1}{6}$.
 The potential for the scalar field is
  \begin{eqnarray}
 V(\phi)=\frac{\kappa^2}{6\alpha}
    \Big[\frac{9 - 54 \alpha}{1 - 2 \alpha}
         +  6 \left( \alpha (7 + 24 \alpha)-2\right) \phi^2
         + \frac{\alpha(1 - 2 \alpha) (3 -16 \alpha) (4-21 \alpha) }
                {1 - 6 \alpha}
                \phi^4
    \Big],
 \end{eqnarray}
which is a $\phi^4$ potential. In the range
$0<\alpha<\frac{1}{6}$, the coefficient of $\phi^4$ term is positive. It is easy to
verify that $V(\phi(y))$ doesn't reach vacuum values when $y$ tends to infinity. From Fig. \ref{figVphiE2A_1},
we know that $\alpha$ can be taken as the parameter that describes the degree of
symmetry breaking: when $\alpha$ becomes smaller, the symmetry is broken more
obviously. Whatever, the parameter $\alpha$ cannot reach $\alpha=0$. In the general
relativity,
 \begin{eqnarray}
 \phi(y)&=&\phi_0 \tanh(\kappa y),\quad
 A(y)=-\frac{\phi_0^2}{9}\left(2\ln \cosh\kappa y+\frac{1}{4}\tanh^2(\kappa y)\right),\\
 V(\phi)&=&\frac{1}{8}\left(\frac{\partial Q}{\partial\phi}\right)^2-\frac{1}{3}Q^2,\quad
 Q=2\kappa\phi_0^2\left(\frac{\phi}{\phi_0}-\frac{1}{3}\frac{\phi^3}{\phi_0^3}\right).
 \end{eqnarray}

It is easy to verify that $F(\phi(y))>0$ is satisfied only if
$\frac{1}{8}\leq\alpha<\frac{1}{6}$. Another kink solution
$\phi=\frac{1}{\sqrt{\alpha}}\tanh(\kappa y)$ was reported in
\cite{DzhunushalievFolomeevMinamitsuji2010}. Numerical information about general
$\alpha$ was provided in \cite{Guo:2011wr}. Non-kink solutions can also lead to the
same warp factor \cite{Bogdanos:2006qw}.

\textbf{Exact solution 2}:
Here, we give another interesting solution:
 \begin{eqnarray}
 F(\phi)&=&\frac{\phi_0^2}{3}+(1-\frac{\phi_0^2}{3})\cosh(\frac{\sqrt{3}\phi}{\phi_0}),\\
 \phi&=&\phi_0 \arctan(\sinh \kappa y),\\
  A(y)&=&-\ln \cosh(\kappa y),\\
  V(\phi)&=&\frac{1}{2} \phi _0^2 \kappa ^2 \cos^2\frac{\phi }{\phi _0}
            +\frac{2 \kappa ^2 }{\sqrt{3}}
             \left(3-\phi_0^2\right)\sin\frac{2\phi }{\phi _0} \sinh\frac{\sqrt{3} \phi }{\phi _0}
               \nonumber\\
          && -\kappa ^2\left(2\phi _0^2
            + (6-2\phi _0^2) \cosh\frac{\sqrt{3} \phi }{\phi _0}\right)
            \sin^2\frac{\phi }{\phi _0}.
\end{eqnarray}

\begin{figure*}[htb]
\begin{center}
\includegraphics[width=4.5cm]{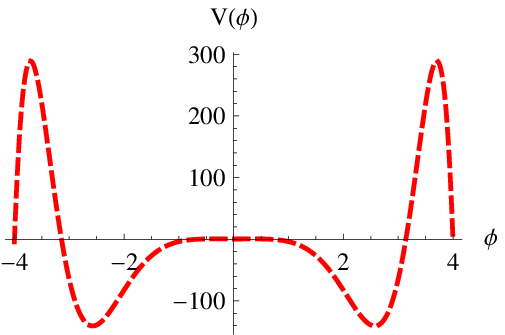}
\includegraphics[width=4.5cm]{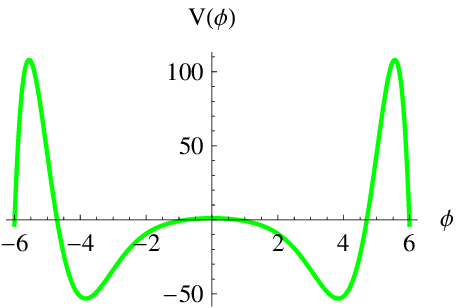}
\includegraphics[width=4.5cm]{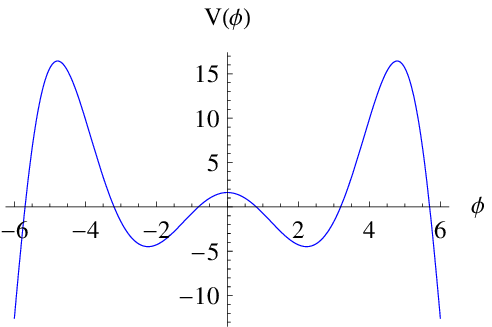}
\includegraphics[width=4.5cm]{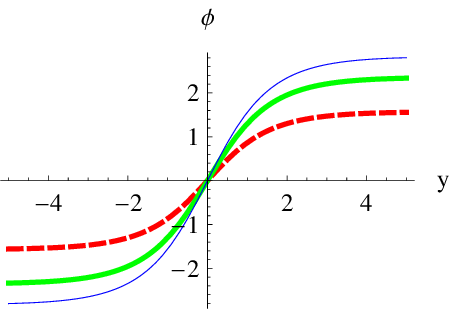}
\includegraphics[width=4.5cm]{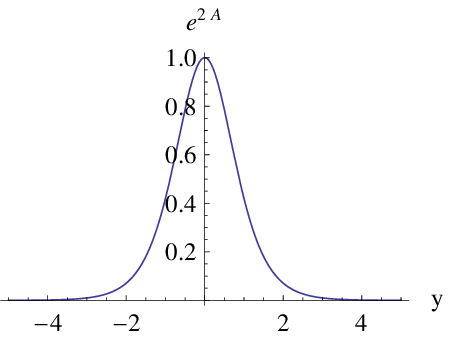}
\end{center}
 \caption{ The shapes of the potential $V(\phi)$, the scalar $\phi$  and the warp factor $e^{2A}$
 for the exact solution 2.
 The dashing red, thick green and thin blue lines correspond to $\phi_0=1, 1.5$ and 1.8, respectively.
 The parameter $\kappa$ is set as $\kappa=1$.}\label{figExact2}
\end{figure*}

In order to guarantee $F>0$, following condition is needed
 \begin{equation}
 0<\phi_0^2<\frac{3\cosh(\sqrt{3}\pi/2)}{3\cosh(\sqrt{3}\pi/2)-1}.
 \end{equation}
 For $\phi_0^2=3$, $F\equiv1$ and $V(\phi)$ is the famous sine-Gordon
potential, which is the case of Einstein's gravity; $\phi_0^2<3$,
$V(\phi)$ oscillatory divergent; $\phi_0^2>3$, $V(\phi)$ negative unlimited. From
Fig. \ref{figExact2}, we know that and $\beta=1-\frac{\phi_0^2}{3}$ can be chosen as
the non-minimal coupling parameter for this solution .

From Fig. \ref{figExact2}, we know that the potential $V(\phi)$ cannot
reach its minimal values when $y$ tends to infinity. For a general solution, we have
 \begin{equation}
 V_\phi(\phi_0)=-10P^2(\phi_0)F_\phi(\phi_0)
 \end{equation}
  for $F(\phi_0)\neq0$. Here
we use $W(\phi_0)=0$ and $\phi_0=\phi(+\infty)$. Obviously in Einstein's gravity,
$V_\phi(\phi_0)=0$, so the scalar reaches vacuum value at infinity for Minkovski
brane. But in non-minimally coupled theories it is not true. For the exact solution 1,
$V_\phi(\phi_0)>0$ since $\alpha>0$. For the exact solution 2, $V_\phi(\phi_0)<0$ if
$\phi_0^2<3$; $V_\phi(\phi_0)>0$ if $\phi_0^2>3$.

\subsection{The deformed solutions}

In this subsection we give two analytic examples for the case of deformed solutions.

 From eq. (\ref{A'}), we know if $(F(\phi))^{1/3}$ is polynomial, the result can be simplified.

  \textbf{Example 1}:
 \begin{eqnarray}
 F(\phi)&=&\left(1-\frac{\phi^2}{\phi^2_0}\right)^3,\\
 \phi&=&\phi_0\tanh (\kappa y),\\
 A(y)&=&-\frac{1}{24} \phi_0^2 \cosh^2(\kappa y)-14 \ln\cosh(\kappa y)+8\kappa y
 \tanh(\kappa y)-\frac{1}{8} \phi_0^2\kappa y \tanh(\kappa y).
 \end{eqnarray}
In this solution, the warp factor descends faster than the one of AdS space-time
 at infinity, and the scalar curvature is divergent.

 \begin{figure*}[htb]
\begin{center}
\includegraphics[width=7cm]{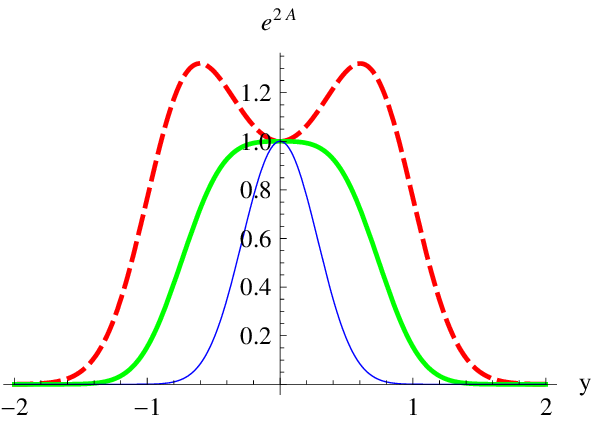}
\includegraphics[width=7cm]{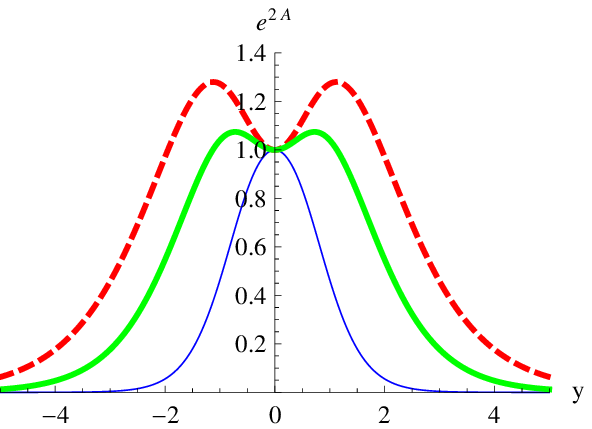}
\end{center}
 \caption{ The shapes of the warp factor $e^{2A}$ for the case of deformed solutions with different $\phi_0$.
In the left figure, corresponding to Example 1, the dashing red, thick green and
thin blue lines correspond to $\phi_0=1, 2.5$ and 5, respectively. In the right one
corresponding to Example 2, the dashing red, thick green and thin blue lines
correspond to $\phi_0=0.5, 1$ and 2, respectively. The parameter $\kappa$ in both
figures is set as $\kappa=1$.}\label{figExmaple}
\end{figure*}

\textbf{Example 2}: In this example, we choose $F(\phi)$ as a periodic function, and
the solution is given by
 \begin{eqnarray}
 F&=&\cos^2(\frac{\phi}{\phi_0}),\\
 \phi&=&\phi_0 \arctan(\kappa y),\\
 A(y)&=& \ln(1+\kappa^2 y^2)-
\frac{(4+\phi_0^2)
 \Big[\text{G}^{23}_{33}\Big( \left.^{1,1,\frac{7}{6}}_{\frac{2}{3},1,0}\right|1+\kappa^2 y^2\Big)
 -\text{G}^{23}_{33}\Big( \left.^{1,1,\frac{7}{6}}_{\frac{2}{3},1,0}\right|1\Big)\Big]}{12
 \Gamma(\frac{2}{3}) \Gamma(\frac{5}{6})},
 \end{eqnarray}
where $G^{23}_{33}$ is the Meijer G function, which is expanded at infinity as
\begin{equation}
\text{G}^{23}_{33}\Big( \left.^{1,1,\frac{7}{6}}_{\frac{2}{3},1,0}\right|1+\kappa^2
y^2\Big)= \frac{1}{\Gamma(\frac{5}{6})}\left[\frac{\pi ^{5/2}(\kappa
y)^{\frac{1}{3}}}{\Gamma(\frac{2}{3})\Gamma(\frac{7}{6})}
-\frac{1}{2}\Gamma(-\frac{1}{6})\ln (\kappa y)\right]+\mathcal
{O}(1),y\rightarrow+\infty.
\end{equation}
So the warp factor is converge. This is consistent with our analysis. The warp
factors of examples 1 and 2 are depicted in Fig. \ref{figExmaple}.

Some deformed solutions were given by numerical calculation in
\cite{Bogdanos:2006qw, Guo:2011wr}. In this subsection, we obtain analytic
solutions. From Fig. \ref{figExmaple}, we know that branes are standard when
$\phi_0$ is small enough; and they become deformed as $\phi_0$ increases. It is a
natural result because of above discussion.

\section{Localization of gravitation }
\label{secStability}

In the braneworld theory, it is an important problem whether the metric
perturbation can be localized on the brane \cite{DeWolfeFreedmanGubserKarch2000,
FarakosKoutsoumbasPasipoularides2007, Bogdanos:2006qw, PhysRevLett.84.2778,
Giddings:2000mu, BronnikovKononogovMelnikov2006, Arnowitt:2005ct, Kehagias:2000au,
Herrera-AguilarMalagon-MorejonMora-Luna2010}. In five-dimensional Minkowski
space-time, the gravitational potential is proportional to $r^{-2}$ according to
Gauss law. However, in general, braneworld models exhibit a four-dimensional
massless graviton which is localized on the brane. So Newtonian potential can be
reproduced.

In this section, we check the stability of the model under the metric tensor
perturbation. It is convenient to consider axial gauge condition:
$h_{5A}=0$\footnote{In other words, vector perturbation vanishes.
However scalar perturbation is kept.}. Under this gauge condition the perturbed metric is
 \begin{equation}
 ds^2=e^{2A}(\eta_{\mu\nu}+h_{\mu\nu})dx^\mu dx^\nu+dy^2.
 \end{equation}
 We can expand modified Einstein equations to first order:
 \begin{eqnarray}\label{fluctuation}
 &&FR^{(1)}_{AB}-e^{2A}
 \left(\frac{2}{3}
    h_{AB}V+\frac{1}{2}h_{AB}'F'+h_{AB}A'F'+
      \frac{1}{3}h_{AB}\nabla^2F(\phi)
      \right)
 -\frac{1}{3}g_{AB}
        \left(2hA'F'+h'F'\right)\nonumber\\
  &&=
  -F_\phi R_{AB}\tilde{\phi}+
  (\nabla_A\nabla_B+\frac{1}{3}g_{AB}\nabla^2)(F_\phi\tilde{\phi})
    +\partial_A\phi\partial_B\tilde{\phi}
       +\partial_A\tilde{\phi}\partial_B\phi
         +\frac{2}{3}g_{AB}V_\phi\tilde{\phi}.
 \end{eqnarray}
Here $R^{(1)}_{AB}$ is the linearized Ricci tensor,
$h=\eta^{\mu\nu}h_{\mu\nu}$,  and $\tilde{\phi}=\delta\phi$.\footnote{In
this section indices $\mu, \nu$ are raised by the flat metric
$\eta^{\mu\nu}$. }

 $R_{AB}^{(1)}$ was calculated in \cite{DeWolfeFreedmanGubserKarch2000}:
\begin{eqnarray}
 R_{\mu\nu}^{(1)}&=&-e^{2A}\left[\frac{1}{2}h_{\mu\nu}''+2A'h_{\mu\nu}'+(A''+4A'^2)h_{\mu\nu}\right]-\frac{1}{2}\Box^{(4)} h_{\mu\nu}\nonumber\\
&&
 -\frac{1}{2}\eta_{\mu\nu}e^{2A}A'h'
     -\frac{1}{2}(\partial_\mu\partial_\nu h-\partial_\mu\partial_\rho h_{\nu}^\rho
     -\partial_\nu\partial_\rho h_\mu^\rho),\\
 R^{(1)}_{55}&=&-\frac{1}{2}(h''+2A'h'),
     \quad    R^{(1)}_{5\mu}=\frac{1}{2}(\partial_\nu h_{\mu}^\nu-\partial_\mu h)',
 \end{eqnarray}
 with $\Box^{(4)}=\eta^{\mu\nu}\partial_\mu\partial_\nu$.

 The fluctuation of the scalar field satisfies
 \begin{equation}
 \nabla^2\tilde{\phi}+\frac{1}{2}\phi'h'+
 \left(\frac{1}{2}F_{\phi\phi}R-V_{\phi\phi}\right)\tilde{\phi}+\frac{1}{2}F_\phi
 R^{(1)}=0.
 \end{equation}
With the help of equations of motion (\ref{Einstein1})-(\ref{Einstein2}), we arrive at
 \begin{eqnarray}
 (\mu,\nu):&&
 F\Big[
   (-\frac{1}{2}h_{\mu\nu}''-2A'h_{\mu\nu}'-\frac{1}{2}\frac{F'}{F}h_{\mu\nu}')
      -\frac{1}{2}e^{-2A}\Box^{(4)} h_{\mu\nu}\nonumber\\
 &&    -\frac{1}{2}\eta_{\mu\nu}A'h'-\frac{1}{3}
        \eta_{\mu\nu}(2h A'+h')\frac{F'}{F}
           -\frac{1}{2}e^{-2A}(\partial_\mu\partial_\nu h
                         -2\partial_{(\nu} h_{\mu)})
    \Big]\nonumber\\
 &=&\eta_{\mu\nu}
  \left[
   F_\phi(A''+4A'^2)\tilde{\phi}+\frac{2}{3}V_\phi\tilde{\phi}
      +\frac{1}{3}\nabla^2(F_\phi\tilde{\phi})
          +A'(F_\phi\tilde{\phi})'
  \right]
   +e^{-2A}F_\phi\partial_\mu\partial_\nu\tilde{\phi},\label{4D fluctuation}\\
 (5,5):&&
  -\frac{F}{2}(h''+2A'h')-\frac{F'}{3}\left(2hA'+h'\right)\nonumber\\
 &=&4F_\phi (A''+A'^2)\tilde{\phi}
      +\frac{2}{3}V_\phi\tilde{\phi}+(F_\phi\tilde{\phi})''
      +\frac{1}{3}\nabla^2(F_\phi\tilde{\phi})+2\phi'\tilde{\phi}',\label{s-fluctuation}\\
 (5,\mu):&&
 \frac{F}{2}\left( h_\mu-\partial_\mu h\right)'
    =\phi'\partial_\mu\tilde{\phi}+(F_\phi\partial_\mu\tilde{\phi})'
       -A'F_\phi\partial_\mu\tilde{\phi}\label{v-fluctuation},
 \end{eqnarray}
 where $h_{\mu}:=\partial^\nu h_{\mu\nu}$.

\subsection{Transverse Traceless component}

First, let us consider the Transverse Traceless ( TT ) component
$\bar{h}_{\mu\nu}$, which is defined by
 \begin{eqnarray}
  \bar{h}_{\mu\nu}:&=&\Pi_{\mu\nu\rho\sigma}h^{\rho\sigma}
  =(\pi_{\mu\rho}\pi_{\nu\sigma}-\frac{1}{3}\pi_{\mu\nu}\pi_{\rho\sigma})h^{\rho\sigma},\\
  h_{\mu\nu}:&=&\bar{h}_{\mu\nu}+h^{NT}_{\mu\nu},
 \end{eqnarray}
with
$\pi_{\mu\nu}=\eta_{\mu\nu}-\partial_\mu\partial_\nu/\Box^{(4)}$.
$\bar{h}_{\mu\nu}$ satisfies
 \begin{equation}
 \partial^\mu \bar{h}_{\mu\nu}=\eta^{\mu\nu}\bar{h}_{\mu\nu}=0.
 \end{equation}
The TT component describes the ordinary gravitational wave and we can prove that TT component is invariant under gauge transformation.
Here we introduce projection operator $\Pi_{\mu\nu\rho\sigma}$, which has following properties:
\begin{eqnarray}
 \Pi_{\mu\nu\rho\sigma}\eta^{\rho\sigma}f_1(x,y)=0,\label{projection1}\\
 \Pi_{\mu\nu}^{~~~\rho\sigma}\partial_{(\rho}w_{\sigma)} (x,y)=0.\label{projection2}
\end{eqnarray}
Here $f_1(x,y), w_\mu(x,y)$ are arbitrary four-dimensional scalar field and vector field, respectively.

From eqs. (\ref{projection1})-(\ref{h_mu}), we can know that the TT component $\bar{h}_{\mu\nu}$ is decoupled with the Non-Transverse Traceless one
$h_{\mu\nu}^{NT}$, and the fluctuation of the background scalar field $\tilde{\phi}$
just refers to the NT component $h_{\mu\nu}^{NT}$. Using the projection operator, We
get fluctuation equation for the TT component \cite{Bogdanos:2006qw}:
 \begin{eqnarray}
 \left
 (-\partial_y^2-(4A(y)+\ln F(\phi(y)))'\partial_y-e^{-2A}\Box^{(4)}
                 \right)\bar{h}_{\mu\nu}=0.  \label{TT fluctuation}
  \end{eqnarray}

  If $F=0$ for some $y$, the stability would have trouble. This coincides with the
  analysis in section \ref{secModel}.

We can eliminate the factor $e^{2A}$ in conformally flat coordinates:
 \begin{equation}
 ds^2=e^{2A(z)}[(\eta_{\mu\nu}+h_{\mu\nu})dx^\mu dx^\nu+dz^2],~~~~dz=
 e^{-A}dy.
 \end{equation}
Now eq. (\ref{TT fluctuation}) yields
 \begin{equation}
 \left(-\partial_{z}^2-3\dot{A}_E(z)\partial_z-\Box^{(4)}\right)\bar{h}_{\mu\nu}=0\label{TT fluctuation2}.
 \end{equation}
Here  $A_E(z)=3A(z)+\ln F(z)$ and the dot denotes the derivative with respect to $z$.

We set $\bar{h}_{\mu\nu}(x,z)=e^{-3A_E/2}e^{ipx}e_{\mu\nu}(p)\varphi(z)$ with
$e_{\mu\nu}$ satisfying $p^\mu e_{\mu\nu}=e^\mu_{~~\mu}=0$. Then eq. (\ref{TT
fluctuation2}) takes Schr\"{o}dinger form:
 \begin{equation}
 \left(-\partial^2_z+V_g(z)\right)\varphi=m^2\varphi,\label{Schrodinger}
 \end{equation}
where $m$ is the four-dimensional mass satisfying $m^2=-p^2$ and the potential for
the gravitons is
 \begin{equation}
 V_g(z)=\frac{3}{2}\ddot{A}_E(z)+\frac{9}{4}\dot{A}_E^2(z).
 \end{equation}
The effective potential for the gravitons is a volcano potential.
 Eq. (\ref{Schrodinger}) can be written as
supersymmetric form
 \begin{eqnarray}
 L^\dag L\varphi(z)&=&m^2\varphi(z), \\
 L&=&\partial_z-\frac{3}{2}\dot{A}_E(z),\\
 L^\dag &=& -\partial_z-\frac{3}{2}\dot{A}_E(z).
 \end{eqnarray}
As the operator $L^\dag L$ is hermitian and positive definite, the solution is
stable under the TT perturbation, or $m^2\geq0$. The massless mode satisfies
$L\varphi_0=0$, and it can be normalized:
 \begin{eqnarray}
 \varphi_0&=&Ne^{3A_E/2}=Ne^{3A(z)/2}F(z)^{1/2},\\
 \int^{+\infty}_{-\infty} dz \varphi_0^2(z)&=&N^2\int
 ^{+\infty}_{-\infty} dz e^{3A(z)}F(z)=N^2\frac{G_5}{ G_4}=1,\\
 N&=&\sqrt{\frac{G_4}{ G_5}}.
 \end{eqnarray}
Here we use the definition of the 4-dimensional Newton's gravity constant.

From eq. (\ref{A'}), we have
 \begin{equation}
 A_E'(y)=-\frac{1}{9}F^{\frac{1}{3}}(\phi(y))
          \int_0^{y}
          \Big[ \frac{4F'^2(\phi(\bar{y}))}
               {3F^{\frac{7}{3}}(\phi(\bar{y}))}
         +\frac{\phi'^2(\bar{y})}
               {F^{\frac{4}{3}}(\phi(\bar{y}))}
    \Big]d\bar{y}.
\end{equation}
Unlike the warp factor, the massless graviton has its maximum value on the brane. It
is clear that the zero mode is localized on the brane, which reproduces the standard
Newtonian gravity on the brane.

\begin{figure*}[htb]
\begin{center}
\includegraphics[width=7cm]{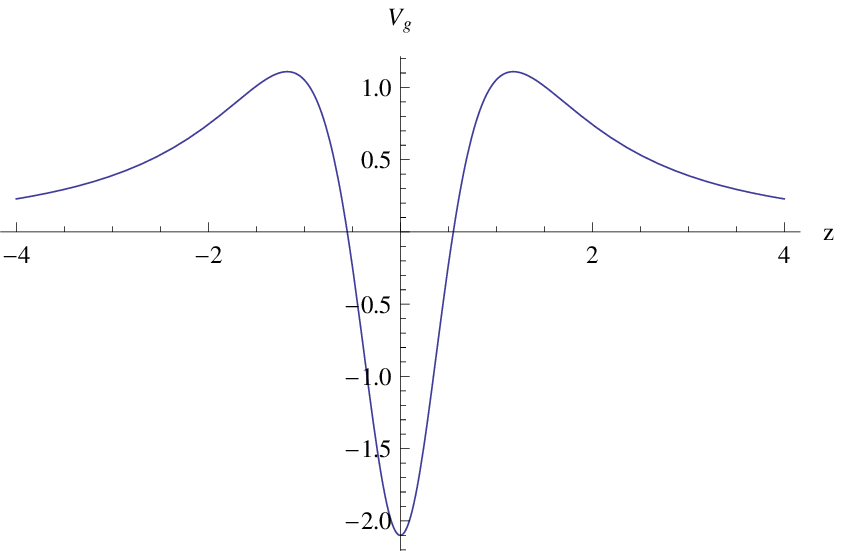}
\includegraphics[width=7cm]{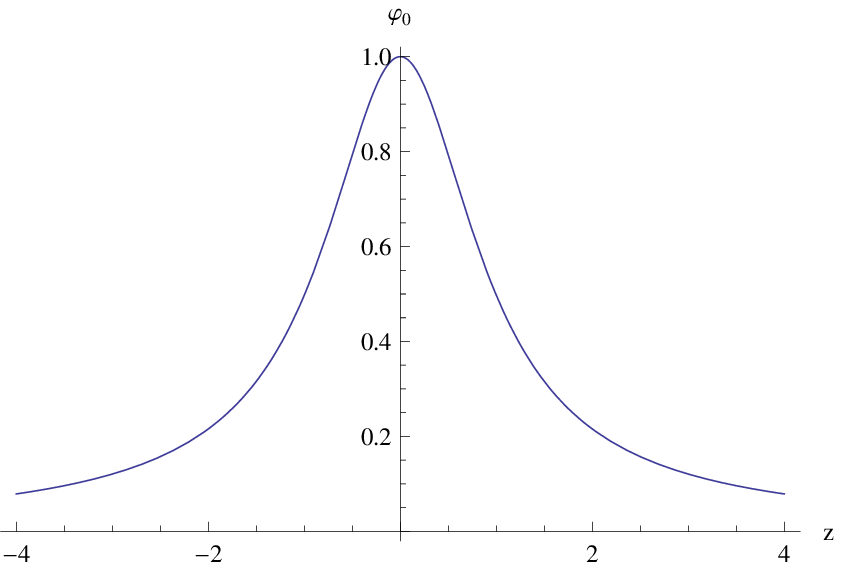}
\end{center}
 \caption{ The shapes of the potential $V_g(\phi)$ and the zero mode
  $\varphi_0$ for the exact solution 1 with $\alpha=1/7$.
 Here $\varphi_0$ is non-normalized. The parameter $\kappa$ is set as
 $\kappa=1$.}\label{figSolution1Vg}
\end{figure*}

We can give analytic expressions of the potential $V_g$ for the exact solution 1
with $\alpha=\frac{1}{7}$ and the exact solution 2. The metrics for both cases share
the same form:
 \begin{eqnarray}
 ds^2&=&\frac{1}{1+(\kappa z)^{2}}(\eta_{\mu\nu}dx^\mu dx^\nu+dz^2).
 \end{eqnarray}
For the exact solution 1 with $\alpha=\frac{1}{7}$, the potential reads
\begin{equation}
V_g(z)=\frac{15 \kappa^2 \left(-14+37 \kappa^2 z^2+28 \kappa^4 z^4+4 \kappa^6
z^6\right)}{4 \left(5+7 \kappa^2 z^2+2 \kappa^4 z^4\right)^2}.
\end{equation}
And the normalized zero mode is
\begin{eqnarray}
    \varphi_{0}(z)
    =\sqrt{\frac{\kappa}{8}}
       \frac{\sqrt{5+2 \kappa^2z^2}}{ \left(1+\kappa^2z^2\right)^{5/4}}.
\end{eqnarray}
We plots the potential $V_g$ and the zero mode $\varphi_0$ for $\alpha=1/7$ in Fig.
\ref{figSolution1Vg}. For other {values} of $\alpha$ in the range of
$\frac{1}{8}\leq\alpha<\frac{1}{6}$, the shape of the potential for the gravitons
does not change sharply.

It is interesting to note that $F(R)$ gravity has the same solution if
$F(R)=R+\gamma R^2$ \cite{LiuZhongZhaoLi2011}, and both solutions have many similar
properties.

 With respect to the exact solution 2, the function $F(\phi(z))$ and the potential
 $V_g(z)$ are
\begin{eqnarray}
F(\phi(z))&=&\frac{\phi_0^2}{3} + (1 - \frac{\phi_0^2}{3}) \cosh(\sqrt{3}
\arctan(\kappa z)),\nonumber\\
 %
 V_g(z)&=&\frac{\kappa^2}{8(1+\kappa^2z^2)^2(\phi_0^2+3\sqrt{1-f^2})^2}
  \Big[45(\phi_0^4-2\phi_0^2+3)\kappa^2z^2-9(\phi_0^4+2\phi_0^2-3) \nonumber\\
  &&+9 \beta (5 \kappa^2 z^2-1) (4\phi_0^2\sqrt{1-f^2}+3\beta+6\beta f^2)
    -{60\sqrt{3}\beta\kappa^3zf (\phi_0^2+3\sqrt{1-f^2})} \Big], ~~~~~~~~
\end{eqnarray}
where
\begin{eqnarray}
f=\sinh(\sqrt{3}\arctan(\kappa z)), \quad \beta=1-\frac{\phi_0^2}{3}.\nonumber
\end{eqnarray}

\begin{figure*}[htb]
\begin{center}
\includegraphics[width=7cm]{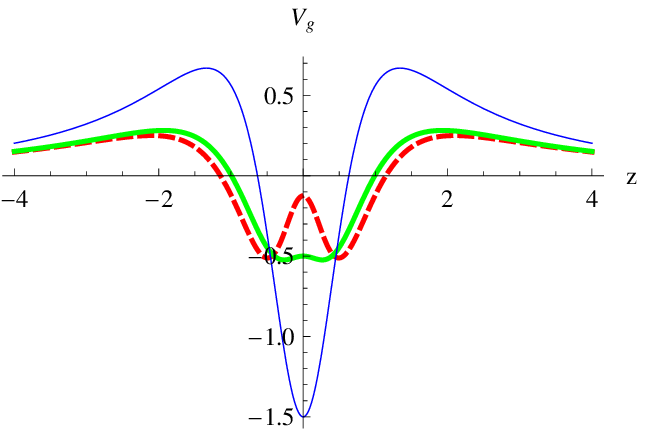}
\includegraphics[width=7cm]{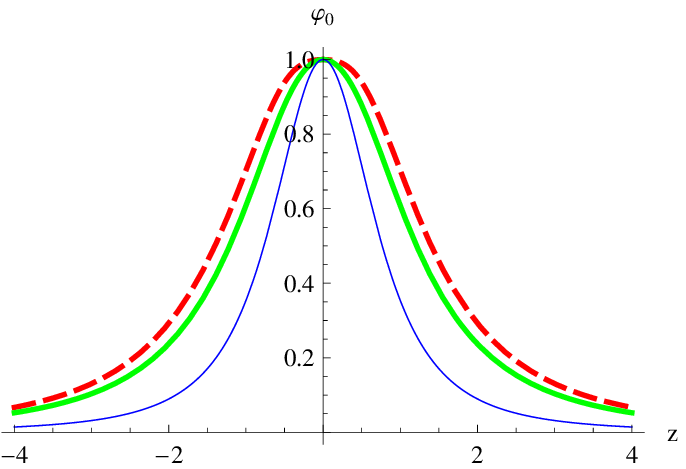}
\end{center}
 \caption{ The shapes of $V_g(\phi), \varphi_0$ for the exact solution 2 with different $\phi_0$.
 Here $\varphi_0$ is non-normalized. The dashing red, thick green and thin blue lines correspond
 to
  $\phi_0=0.5, 1$ and $\sqrt{3}$,
respectively. The parameter $\kappa$ is set as $\kappa=1$.}\label{figSolution2Vg}
\end{figure*}

We find that when $\phi_0^2<\frac{23 - \sqrt{337}}{4}$, the potential $V_g$ is a
double-well volcano potential; when $\phi_0^2>\frac{23 - \sqrt{337}}{4}$, we get a
single-well volcano potential. This point is interpreted in Fig.
\ref{figSolution2Vg}.

In this solution, the property of the potential for the gravitons mainly depends on
$F(\phi)$, or more precisely, on $\phi_0$, since in this solution the warp factor is
fixed.

\begin{figure*}[htb]
\begin{center}
\includegraphics[width=7cm]{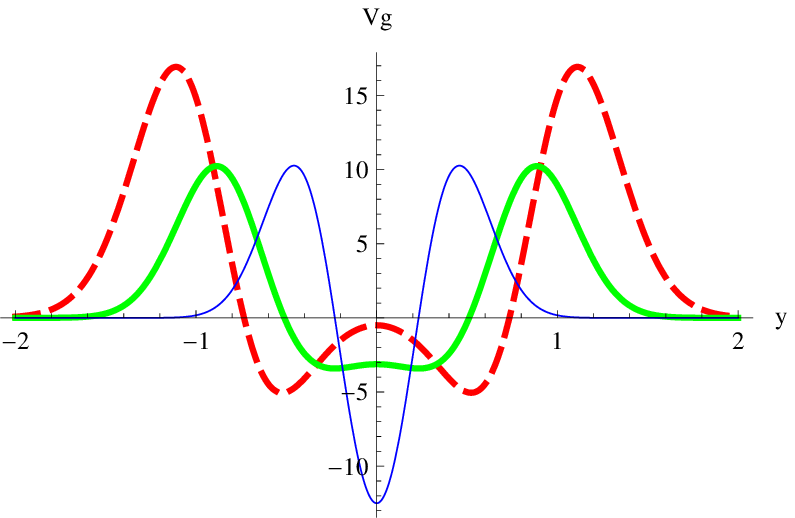}
\includegraphics[width=7cm]{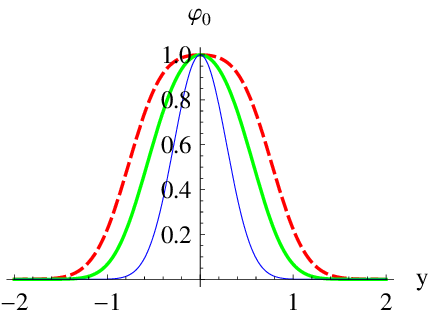}
\end{center}
 \caption{ The shapes of $V_g(y)$ and $\varphi_0$ for the example 1 with different $\phi_0$.
The dashing red, thick green and thin blue lines correspond to
  $\phi_0=1, 2.5$ and $5$,
respectively. The parameter $\kappa$ is set as $\kappa=1$.}\label{figsolutionVg}
\end{figure*}

For the deformed solutions, since the non-minimal coupling function $F(\phi)$ and
the warp factor $e^{2A}$ have similar properties in the examples 1 and 2, we only
plot the potential $V_g$ for the example 1 in Fig. \ref{figsolutionVg}.

\subsection{Non-Transverse Traceless component}

Second, we consider the stability for the Non-Transverse-Traceless (NT) part.

The NT part $h^{NT}_{\mu\nu}$ is decoupled with the TT component, and is
only dependent on $h_\mu$ and $h$:
\begin{eqnarray}
 h_{\mu\nu}^{NT}&=&2\partial_{(\mu}\left(\frac{1}{\Box^{(4)}}h_{\nu)}\right)
    +\frac{1}{3}\eta_{\mu\nu}(h-\partial_\rho h^\rho)
    +\partial_\mu \partial_\nu
    \left[
    \frac{1}{3\Box^{(4)}}(\frac{2}{\Box^{(4)}}\partial_\rho h^\rho+h)
    \right],~~~~\label{NT def}\\
 \eta^{\mu\rho}\partial_\rho h^{NT}_{\mu\nu}&=& h_{\nu},\quad
 \eta^{\mu\nu}h_{\mu\nu}^{NT}=h.
\end{eqnarray}

From eq. (\ref{v-fluctuation}), we know there exists a function
$f(x,y)$ such that
 \begin{equation}
 h_{\mu}(x,y)= \partial_\mu f(x,y).\label{h_mu}
 \end{equation}

From eq. (\ref{h_mu}), we know that eq. (\ref{NT def}) can
be rewritten as the following form:
\begin{equation}
h^{NT}_{\mu\nu}(x,y)=2\partial_\mu \partial_\nu \Phi(x,y)+2\eta_{\mu\nu}\Psi(x,y).
\end{equation}
Now we consider the following gauge transformation:
 \begin{eqnarray}
  x^\mu\rightarrow x^\mu- \xi^\mu(x,y),\quad
  y\rightarrow y-\xi^5(x,y).
 \end{eqnarray}
The corresponding gauge transformation for the perturbation $h_{AB}$ is: \footnote{In this subsection $h_{\mu\nu}^{NT}$ is written as $h_{\mu\nu}$ for short since the TT component is gauge invariant.}
\begin{eqnarray}
 &&ds^2=e^{2A}(\eta_{\mu\nu}+h_{\mu\nu})dx^\mu dx^\nu
      +2h_{5\nu}dydx^\nu +(1+h_{55})dy^2,\\
 &&h_{\mu\nu}\rightarrow h_{\mu\nu}+
          2\partial_{(\mu }\xi_{\nu)} +2\eta_{\mu\nu}\xi^5,\\
 &&h_{5\nu}\rightarrow h_{5\nu}+ \partial_\nu \xi^5+e^{2A}\xi_\nu', \\
 &&h_{55}\rightarrow  h_{55}+ 2(\xi^5)'.
\end{eqnarray}

We choose the proper functions $\xi^\mu$ and $\xi^5$ to eliminate the
term such as $\partial_\mu\partial_\nu\Phi$:
\begin{eqnarray}
 \xi_\mu&=&-\partial_\mu \Phi,\\
 \xi^5&=&e^{2A}\Phi'.
\end{eqnarray}
This choice can keep the gauge condition $h_{5\mu}=0$, however $h_{55}=0$ cannot be
preserved at the same time.

 Then the NT perturbation can be written in the familiar
 form:\footnote{Fluctuation in this form is usually called scalar
 perturbation.}
 \begin{eqnarray}
 ds^2=e^{2A}\left[(1+2\Psi+2e^{A}\dot{\Phi})\eta_{\mu\nu}dx^\mu dx^\nu+(1+2\ddot{\Phi}+2\dot{A}\dot{\Phi})dz^2\right].
 \end{eqnarray}

To simplify our discussion, we make a conformal transformation
(\ref{Einstein_frame_metric}) (\ref{transformation}) to recover Einstein frame:
\begin{eqnarray}
 \tilde{\xi}&=&\tilde{\phi}\sqrt{\frac{F_\phi^2}{3F^{2}}+\frac{1}{4F}},\\
 ds^2_E&=&e^{2A_E}\left[(1+2\vartheta)\eta_{\mu\nu}dx^\mu dx^\nu+(1+2\zeta)dz^2\right],\\
 \vartheta:&=&\Psi+e^{A}\dot{\Phi}+\frac{F_\phi}{3F}\tilde{\phi},
 \quad\zeta:=\ddot{\Phi}+\dot{A}\dot{\Phi}+\frac{F_\phi}{3F}\tilde{\phi}.
\end{eqnarray}
 Following Ref. \cite{Kobayashi:2001jd}, we have
 \begin{eqnarray}
(\mu,\nu): &&\left(
  3\ddot{\zeta}-6\ddot{A}_E\vartheta-3\dot{A}_E\dot{\vartheta}+9\dot{A}_E\dot{\zeta}-6\dot{A}_E^2\vartheta+
   \Box^{(4)}(\vartheta+2\zeta)\right)\eta_{\mu\nu}
                   \nonumber\\
  &&-\partial_\mu\partial_\nu(\vartheta+2\zeta)=
  (-\dot{\xi}\dot{\tilde{\xi}}+\vartheta\dot{\xi}^2
     -e^{2A_E}\frac{\partial\tilde{V}}{\partial\xi}\tilde{\xi})\eta_{\mu\nu},\label{scalar_munu}\\
 (\mu,5):&&3\partial_\mu(-\dot{\zeta}+\dot{A}_E\vartheta)=\dot{\xi}\partial_\mu\tilde{\xi},\label{scalar_mu5}\\
 (5,5):&&3\Box^{(4)}\zeta+12\dot{A}_E\dot{\zeta}-12\dot{A}_E^2\vartheta=
         \dot{\xi}\dot{\tilde{\xi}}-\vartheta\dot{\xi}^2
         -e^{2A_E}\frac{\partial\tilde{V}}{\partial\xi}\tilde{\xi}.\label{scalar_55}
 \end{eqnarray}
The perturbation equation for the scalar field is
\begin{equation}
\ddot{\tilde{\xi}}+3\dot{A}_E\dot{\zeta}+(4\dot{\zeta}-\dot{\vartheta}-6\dot{A}_E\vartheta)\dot{\xi}-2\vartheta\xi''+\Box^{(4)}\tilde{\xi}
      =e^{2A_E}\frac{\partial^2\tilde{V}}{\partial\xi^2}\tilde{\xi}.
\end{equation}

Noticing the off-diagonal part of eq.(\ref{scalar_munu}), we get
\begin{equation}
\vartheta+2\zeta=0\label{off-diag}.
\end{equation}
Solving eq.(\ref{scalar_mu5}), we get
\begin{equation}
\tilde{\xi}=\frac{3}{\dot{\xi}}(\dot{A}_E\vartheta-\dot{\zeta}).\label{scalar}
\end{equation}
Substituting the relations (\ref{off-diag})-(\ref{scalar}) into eq.
(\ref{scalar_munu})+(\ref{scalar_55}), we have
\begin{equation}
 \Box^{(4)}\zeta+\ddot{\zeta}+
   9\dot{A}_E\dot{\zeta}+(4\ddot{A}_E+12\dot{A}_E^2)\zeta
   =e^{2A_E}\frac{\partial\tilde{V}}{\partial\xi}\frac{2}{\dot{\xi}}(2\dot{A}_E\zeta+\dot{\zeta}).
   \label{scalar-b}
\end{equation}
Noticing the EOM of the background scalar:
 \begin{equation}
  \ddot{\xi}+3\dot{A}_E\dot{\xi}=e^{2A_E}\frac{\partial\tilde{V}}{\partial\xi},
 \end{equation}
 eq. (\ref{scalar-b}) can be rewritten as
 \begin{equation}
\Box^{(4)}\zeta+\ddot{\zeta}+
   (3\dot{A}_E-2\frac{\ddot{\xi}}{\dot{\xi}})\dot{\zeta}+
   4(\ddot{A}_E-\dot{A}_E\frac{\ddot{\xi}}{\dot{\xi}})\zeta=0.
\end{equation}
Separating the coordinates $\zeta=\dot{\xi}e^{-3A_E}S(z)e^{ipx}$, we obtain a
Schr\"{o}dinger-type equation
 \begin{equation}
 (-\partial_z^2+V_N(z))S(z)=m^2S(z),
 \end{equation}
where $m^2=-p^2$ and the effective potential $V_N$ for the NT perturbation is
\begin{equation}
V_N=-\frac{5}{2}\ddot{A}_E+\frac{9}{4}\dot{A}_E^2+\dot{A}_E\frac{\ddot{\xi}}{\dot{\xi}}
    -\frac{\partial_z^3\xi}{\partial_z\xi}+2\left(\frac{\ddot{\xi}}{\dot{\xi}}\right)^2.
\end{equation}
If the transformation (\ref{transformation}) is regular, in other words,
$F(\phi(\infty))>0$, the potential $V_N\rightarrow
0^{+}$ as $y\rightarrow \infty$. If $m^2<0$, $S(z)$ would diverge either at $y=+\infty$ or at $y=-\infty$. So
there is also no tachyon mode in the NT perturbation. Furthermore, we guess that there
should not exist bound state under the NT perturbation. Otherwise, the Newtonian
potential would be changed. For $F(\phi(\infty))=0$, $\xi$ is not a kink scalar, so
the solution is unstable.

From above discussion, we learn that the stability for the NT perturbation needs
additional restrictions. Noticing that $\tilde{\phi}$ vanishes for the TT part, we don't
care about the stability of the background scalar. The stability of the background
scalar plays a vital role in proving the results in this subsection.

Finally, we conclude that this model is stable under the perturbation if
the space-time is asymptotic AdS.

 \section{Conclusion}

In this paper, we have studied thick Poincar\'{e} brane scenario with a non-minimally coupled bulk scalar. In our
model, the gravitational action is modified by $F(\phi)R$. We list our results as follows:

\begin{itemize}
\item{There are two categories of the brane configuration: standard branes and deformed ones. The
former is just the same with the solutions in general relativity, whereas the warp factor can reach its peak beyond the brane in the latter.
}

\item{The asymptotic space-time in some cases may be not $AdS_5$.
We conjecture that the warp factor $e^{2A}$ vanishes at
infinity in our model. The asymptotic space-time can be Minkowski only in the
sense that the Ricci curvature is zero.}

\item{This model is stable under the Transverse Traceless perturbation.
The spectrum for the TT component contains a massless mode (the 4D massless
graviton) and a tower of continuous gapless massive KK modes. The massless graviton
can be normalized and localized on the brane.}

\item{This model is stable under the Non-Transverse Traceless perturbation only if asymptotic space-time is AdS.
}
\end{itemize}

In a complete braneworld model, the modified Newtonian potential should be given.
For a non-minimal coupling gravity, it means that the four-dimensional Brans-Dicke
parameter is required. In \cite{FarakosKoutsoumbasPasipoularides2007} the
Brans-Dicke parameter was calculated for RS metric. However, the detailed
calculation is not easy for thick brane solutions because the
technic used in RS metric is not valid. Another important problem is
that there should not exist bound state under the NT perturbation, however we cannot
finish the proof. We expect some related works can be reported in the near future.

\section*{Acknowledgments}

This work was
supported by the Program for New Century Excellent Talents in
University, the Huo Ying-Dong Education Foundation of Chinese
Ministry of Education (No. 121106), the National Natural Science
Foundation of China (No. 11075065),  and the Fundamental Research Funds for the Central Universities (No. lzujbky-2012-k30).


\end{document}